\documentclass[english,12pt]{article}       

\usepackage{graphicx,amsmath,amssymb,color}
\usepackage{mathptmx}   
\usepackage[T1]{fontenc}
\usepackage[utf8]{inputenc}

\newcommand{\cn}{\,{\sf cn}}
\newcommand{\sn}{\,{\sf sn}}
\newcommand{\dn}{\,{\sf dn}}
\newcommand{\sech}{\,{\sf sech}}

\begin{document}

\title{New exact superposition solutions to KdV2 equation
}

\author{Piotr Rozmej$\;^1$ and Anna Karczewska$\;^2$ 
\\ \small
$^1$  Faculty of Physics and Astronomy,\\  \small University of Zielona G\'ora, Szafrana 4a, 65-246 Zielona G\'ora, Poland \\ \small
$^2$ Faculty of Mathematics, Computer Science and Econometrics,\\ \small University of Zielona G\'ora, Szafrana 4a, 65-246 Zielona G\'ora, Poland
}



\date{}
\maketitle

\begin{abstract}
New exact solutions to the KdV2 equation (known also as the extended KdV equation) are constructed. The KdV2 equation is a second order approximation of the set of Boussinesq's equations for shallow water waves which in first order approximation yields KdV. The 
exact solutions 
~$\frac{A}{2}\left(\dn^2[B(x-vt),m]\pm \sqrt{m}\,\cn [B(x-vt),m]\dn [B(x-vt),m]\right)+D$~ in the form of periodic functions 
found in the paper complement other forms of exact solutions to KdV2 obtained earlier, i.e., the solitonic ones 
and periodic ones given by a single $\cn^2$ or $\dn^2$ Jacobi elliptic functions.
\end{abstract}



\section{Introduction} \label{intro}

Water waves have attracted the interest of scientists for at least two centuries. One hundred and seventy years ago Stokes \cite{Stokes}  showed that waves described by nonlinear models can be periodic. In this way he pioneered the field of nonlinear hydrodynamics. The next important step was made by Boussinesq
\cite{Bouss}, though his achievement went unnoticed for many years. 
The most important approximation of the set of Euler's hydrodynamic equations was made by Korteweg and de Vries \cite{KdV} who obtained a single nonlinear dispersive wave equation, called nowadays the KdV equation in their names.
KdV became so famous since it constitutes a first order approximation for nonlinear waves in many fields: hydrodynamics, magneto-hydrodynamics, electrodynamics, optics, mathematical biology, see, e.g., monographs \cite{DrJ,InRo,Rem}. It consists of the mathematically simplest terms representing the interplay of nonlinearity and dispersion. 
For some ranges of values of equation coefficients these two counteracting effects may cancel admitting solutions in the form of unidirectional waves of permanent shapes.
KdV is integrable and possesses an infinite number of invariants. Its analytic solutions were found in the forms of single solitons, multi-solitons and periodic (cnoidal functions), see, e.g. \cite{GGKM,ZK} and monographs \cite{DrJ,InRo,Rem,AbC,Whit,Ding}.

For the shallow water problem leading to KdV, two small parameters are assumed: wave amplitude/depth $\alpha=(A/H)$ and depth/wavelength squared $\beta=(H/L)^2$. Then the perturbation approach to Euler's equations for the  irrotational motion of inviscid fluid is applied. Limitation to the terms of first order in $\alpha,\beta$, yields the KdV equation in the following form (expressed in scaled dimensionless variables in fixed reference frame)
\begin{align} \label{kdv}
\eta_t & +  \eta_x + \frac{3}{2} \alpha\,\eta\eta_x+ \frac{1}{6}\beta\, \eta_{3x}  =0 . 
\end{align}
Here and below the low indexes denote partial derivatives, e.g., $\eta_t\equiv \frac{\partial \eta}{\partial t} \mbox{~and~}  \eta_{kx}\equiv \frac{\partial^k \eta}{\partial x_k}$.

One of reasons for the enormous success of the KdV equation is its simplicity and integrability. However, KdV is derived under the assumption that both $\alpha$ and $\beta$ parameters are small. Therefore one should not expect that KdV can properly describe shallow water waves for larger values of parameters $ \alpha,\beta$. In principle, the extended KdV equation, obtained in second order approximation with respect to these parameters, should be applicable for a wider range of parameters $ \alpha,\beta$ than the KdV equation.

The next, second order approximation to Euler's equations for long waves over a shallow riverbed is 
\begin{align} \label{kdv2}
\eta_t & +  \eta_x + \frac{3}{2} \alpha\,\eta\eta_x+ \frac{1}{6}\beta\, \eta_{3x} -\frac{3}{8}\alpha^2\eta^2\eta_x 
 + \alpha\beta\,\left(\frac{23}{24}\eta_x\eta_{2x}+\frac{5}{12}\eta\eta_{3x} \right)+\frac{19}{360}\beta^2\eta_{5x} =0 . 
\end{align}
This equation was first derived by Marchant and Smyth \cite{MS90} and called the {\tt extended KdV}. Later (\ref{kdv2})  was derived in a different way in \cite{BS13} and as a by-product in derivation of the equation for waves over uneven bottom in \cite{KRR14,KRI14}. Therefore exact solutions of (\ref{kdv2}) are the best initial conditions for performing numerical evolution of waves entering the regions where the bottom changes, see \cite{RRIK}. We call it {\bf KdV2}.
KdV2 is not integrable. Contrary to KdV it has only one conservation law (mass or equivalent volume). On the other hand, as pointed out in \cite{KRIR}, there exist adiabatic invariants of KdV2 in which relative deviations from constant values are very small (of the order of $0(\alpha^3$)).

Despite its nonintegrability, KdV2 possesses exact analytic solutions. The first class of such solutions, single solitonic solutions found  by us is published 
in \cite{KRI14}. These solutions have the same form $\sech^2$ as KdV solitonic solutions but slightly different coefficients. Following this discovery we came across  papers by Khare and Saxena \cite{KhSa,KhSa14,KhSa15} who showed that there exist classes of nonlinear equations which possess exact solutions in the form of hyperbolic or Jacobi elliptic functions. They showed also that besides the usual solitonic ($\sech^2$) and periodic cnoidal ($\cn^2$) solutions other new solutions in the form of superpositions of hyperbolic or Jacobi elliptic functions exist.
Since KdV belongs to these classes we formulated the hypothesis that this applied to KdV2, as well. In \cite{IKRR17} this hypothesis was verified for periodic cnoidal ($\cn^2$) solutions. Using an algebraic approach we showed there that for KdV2 (\ref{kdv2}) there exist analytic periodic solutions in the same form as KdV solutions, that is,
\begin{equation} \label{cn2}
\eta(x,t) = \cn^2[B(x-vt),m]+D.
\end{equation}
The formulas for the parameters of solutions $B,D,v,m$ were given explicitly as functions of the coefficients $ \alpha,\beta$ of KdV2 (\ref{kdv2}). It appeared that KdV2 imposed some restrictions on the ranges of solution parameters when compared to KdV. 

In \cite{RKI}, following Khare and Saxena \cite{KhSa}, we checked our hypothesis for KdV2 solutions of the form 
\begin{align} \label{etapm}
 \eta_{\pm}(y) & =  \frac{A}{2} \left\{ \dn^2[B(x-vt),m] 
\pm\sqrt{m}\cn(By,m)\dn[B(x-vt),m] \right\}. 
\end{align}
It was proved that both functions  (\ref{etapm}) satisfy (\ref{kdv}), moreover explicit formulas for coefficients of (\ref{etapm}) were given. However, this closed-form mathematical solution does not fulfil important physical condition that the mean fluid level remains the same for arbitrary wave.

This article complements \cite{KRI14,IKRR17,RKI} by giving physically relevant solutions to KdV2 in the form of superpositions. In section \ref{sss2} equations determining the coefficients of the superposition solution are derived. The conditions necessary for the solution to describe shallow water waves are imposed on the solutions in section \ref{sss3} and formulas for the coefficients of the solution are obtained. In section \ref{sss4} some  examples of solutions are presented with additional verification by numerical evolution of the obtained solutions according to (\ref{kdv2}). Section \ref{concl} contains 
 conclusions.

 Our idea to look for exact solutions to KdV2 in the same forms as solutions to KdV gained recently a strong support by results of Abraham-Shrauner \cite{BAS18}.

\subsection{Algebraic approach to KdV}
In order to present the approach we show the KdV case first.
Assume solutions to KdV 
in the following form
\begin{equation} \label{eypm}
 \eta_{\pm}(y) = \frac{A}{2} \left[ \dn^2(By,m) \pm\sqrt{m}\cn(By,m)\dn(By,m) \right] +D,
\end{equation}
where $A,B,D,v$ are yet unknown constants ($m$ is the elliptic parameter) which have the same meaning as in a single $\cn^2$ solution. 
Coefficient $D$ is necessary in order to maintain, for arbitrary $m$, the same volume for a wave's elevations and depressions with respect to the undisturbed water level.

Introduce $y:=x-v t$. Then 
 $\eta(x,t) =\eta(y)$,  $\eta_t=-v\eta_y$~ and equation (\ref{kdv}) takes the form of an ODE
\begin{align} \label{kdvy}
(1-v) \eta_y & + \frac{3}{2} \alpha\,\eta\eta_y+ \frac{1}{6}\beta\, \eta_{3y} =0.
\end{align}
Insertion of (\ref{eypm}) into (\ref{kdvy}) gives (common factor $ CF = AB \sqrt{m}\left(\sqrt{m} \cn +\dn \right)^2 \sn)$)
\begin{align} \label{0ry} 
CF\,\left(F_0+F_2 \cn^2+ F_{11} \cn\dn \right) & =  0.
\end{align}
Then there are three conditions on the solution
\begin{align} \label{f0}  
F_0 & =9 \alpha  A-9 \alpha  A m-2 \beta  B^2+10 \beta  B^2 m+18 \alpha 
   D-12 v+12 =0, \\  \label{f2}
F_2 & =9 \alpha  A m-12 \beta  B^2 m  = 0, \\ \label{f11}
 F_{11} & = 9 \alpha  A \sqrt{m}-12 \beta  B^2 \sqrt{m} = 0.
\end{align}
Equations (\ref{f2}) and (\ref{f11})  are equivalent and yield
\begin{align} \label{f2-11} 
B & = \sqrt{ \frac{3\alpha}{4\beta}\, A}.
\end{align}
Insertion this into (\ref{f0}) gives
\begin{align} \label{f000} 
\alpha  A (m-5)-4 (3 \alpha  D+2)+8 v & =  0,
\end{align}
Periodicity condition 
implies
\begin{align} \label{B000} 
L= & \frac{4 K(m)}{B}.
\end{align}
Then volume conservation condition
determines $D$ as
\begin{align} \label{d000} 
D= & -\frac{A}{2} \frac{E(m)}{K(m)} .
\end{align}
Finally insertion of $D$ into (\ref{f000}) gives velocity as
\begin{align} \label{abdV} 
v & = 1+ \frac{\alpha A}{8} \left[ 5-m-6\, \frac{E(m)}{K(m)}\right]\equiv 1+ \frac{\alpha A}{8}\,EK(m).
\end{align}
$K(m)$ and $E(m)$ which appear in (\ref{B000})-(\ref{abdV}) are the complete elliptic integrals of the first kind and the second kind, respectively.

Equations (\ref{f2-11}), (\ref{d000}) and (\ref{abdV}) express coefficients $B,D,v$  of the superposition solution   (\ref{etapm}) as functions of the amplitude $A$, elliptic parameter $m\in [0,1]$ and parameters $\alpha,\beta$ of the KdV equation. In principle, these equations admit arbitrary amplitude of KdV solution in the form (\ref{etapm}). 

\begin{figure}[tbh]
\begin{center} 
\resizebox{0.7\columnwidth}{!}{\includegraphics{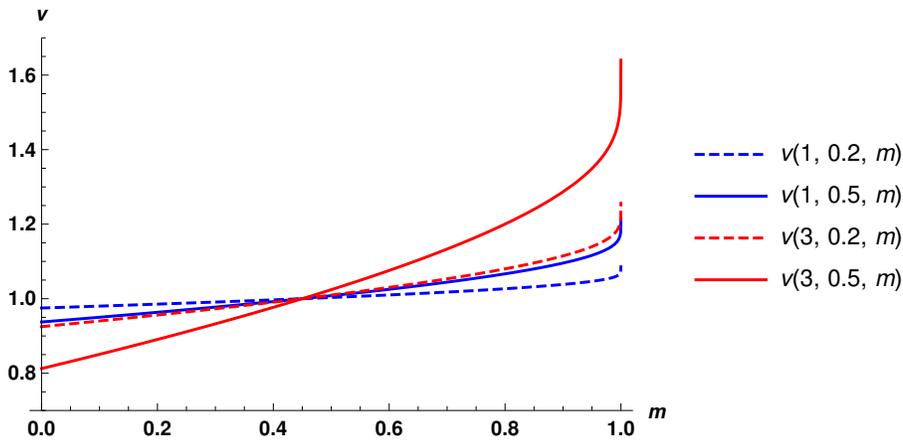}}
\end{center}\vspace{-3mm}
\caption{Velocity $v(A,\alpha,m)$ (\ref{abdV}) of the solution (\ref{etapm}) for different $A$ and $\alpha$.}\label{vS+kdv}
\end{figure}

Coefficients $B,D,v$ and the wavelength $L$ obtained above for solution (\ref{etapm}) are different form coefficients of usual cnoidal solutions in the form (\ref{cn2}). 
\begin{figure}[tbh]
\begin{center} 
\resizebox{0.5\columnwidth}{!}{\includegraphics{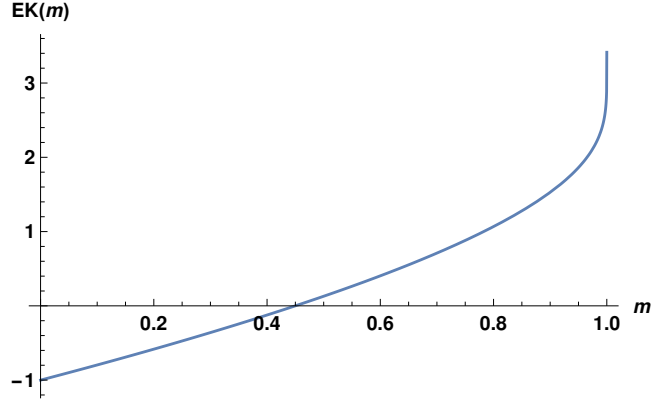}}
\end{center}\vspace{-3mm}
\caption{Function $EK(m)$ given by (\ref{EKm}). }\label{fAK}
\end{figure}
In particular, since the function 
\begin{equation} \label{EKm}
EK(m)=\left[ 5-m-6\, \frac{E(m)}{K(m)}\right]
\end{equation}
changes its sign at $m\approx 0.449834$ the velocity dependence of the wave (\ref{etapm}) is much different than that of the $\cn^2$ wave [For $\cn^2$ wave $v=1+\frac{\alpha A}{2m}\left(2-m -3\,\frac{E(m)}{K(m)} \right) $], see Eq.~(24) in \cite{IKRR17}.  
Examples of $m$ dependence of the velocity (\ref{abdV}) are displayed in figure \ref{vS+kdv}.

\section{Algebraic approach to KdV2} \label{sss2}

Now, we look for  solutions to KdV2 (\ref{kdv2}) in the same form (\ref{eypm}). In this case the corresponding ODE takes the form 
\begin{align} \label{kdv2y}
(1-v) \eta_y & + \frac{3}{2} \alpha\,\eta\eta_y+ \frac{1}{6}\beta\, \eta_{3y} -\frac{3}{8}\alpha^2\eta^2\eta_y  
 + \alpha\beta\,\left(\frac{23}{24}\eta_y\eta_{2y}+\frac{5}{12}\eta\eta_{3y} \right)+\frac{19}{360}\beta^2\eta_{5y} = 0 . 
\end{align}

Assume solutions to KdV2 in the form (\ref{eypm}),
$$ 
 \eta_{\pm}(y) = \frac{A}{2} \left[ \dn^2(By,m) \pm\sqrt{m}\cn(By,m)\dn(By,m) \right] +D,
$$ 
where $A,B,D,v$ are yet unknown constants ($m$ is the elliptic parameter) which have the same meaning as previously.

Insertion of (\ref{eypm}) to (\ref{kdv2y}) yields
\begin{align} \label{ry} CF 
\left(F_0+F_2 \cn^2+F_4 \cn^4+ F_{11}\cn\dn +F_{31} \cn^3\dn \right) & =  0,
\end{align}
where common factor is 
$$ CF = AB \sqrt{m}\left(\sqrt{m} \cn +\dn \right)^2 \sn .$$
Equation (\ref{ry}) is satisfied for arbitrary arguments when all coefficients $F_0,\ldots,F_{31}$ vanish simultaneously. This imposes five  conditions on parameters 
\begin{align} \label{ry0}
F_0& =  135 \alpha ^2 A^2 (m-1)^2+60 \alpha  A (m-1) \left(\beta  B^2 (48
   m-5)-9 \alpha D+18\right)\nonumber \\ & \hspace{3ex}
-4 \left[19 \beta ^2 B^4
   \left(61 m^2-46 m+1\right)+30 \alpha D \left(5 \beta 
   B^2 (5 m-1)+18\right)  \right. \\ & \left. \hspace{3ex}
+60 \beta  B^2 (5 m-1)-135 \alpha ^2
  D^2+360\right]+1440 v = 0 , \nonumber
\\ \label{ry2}
F_2& =  -15 m \left[27 \alpha ^2 A^2 (m-1)-12 \alpha  A \left(\beta  B^2
   (37-59 m)+3 \alpha D-6\right)\right.  \\ &\left.\hspace{7ex}
-16 \beta  B^2 \left(38 \beta  B^2 (2 m-1)+15 \alpha D+6\right)\right] =0,\nonumber
\\ \label{ry4}
F_4& =  90 m^2 \left(3 \alpha ^2 A^2+86 \alpha  A \beta  B^2-152 \beta ^2
   B^4\right) = 0 ,
\\ \label{ry11}
F_{11}& =  -30 \sqrt{m} \left[9 \alpha ^2 A^2 (m-1)-3 \alpha  A \left(\beta 
   B^2 (31-75 m)+6 \alpha D-12\right)\right.  \\ &\left.\hspace{10ex}
-4 \beta  B^2
   \left(19 \beta  B^2 (5 m-1)+30 \alpha 
  D+12\right)\right] = 0 ,\nonumber
\\ \label{ry31}
F_{31}& =  90 m^{3/2} \left(3 \alpha ^2 A^2+86 \alpha  A \beta  B^2-152 \beta
   ^2 B^4\right) = 0 .
\end{align}

As previously in \cite{RKI} equations (\ref{ry4}) and (\ref{ry31}) are equivalent. 
Denote ~$\displaystyle z=\frac{\beta B^2}{\alpha A}$. 
Then  roots of (\ref{ry31}) are the same as for solitonic \cite{KRI14} and $\cn^2$ solutions \cite{IKRR17}, that is 
\begin{equation} \label{z_12}
z_{1/2} = \frac{43\mp \sqrt{2305}}{152}.
\end{equation}
In principle we should discuss both cases. 

Express equations (\ref{ry0}), (\ref{ry2}) and (\ref{ry11}) through $z$
by substituting
\begin{equation} \label{BB2}
B =\sqrt{\frac{A\alpha\, z}{\beta}}.
\end{equation}
This gives 
\begin{align} \label{R2}
608 \alpha  A (2 m-1) z^2 &+12 z (\alpha  A (37-59 m)+20
   \alpha  D+8)
-9 (3 \alpha  A (m-1)-4 \alpha  D+8)=0
\end{align} 
and
\begin{equation} \label{R11}
9 A (\alpha -\alpha  m)+76 \alpha  A (5 m-1) z^2+3 \alpha  A (31-75 m) z+18
   \alpha D+24 z (5 \alpha D+2)-36 =0
\end{equation}
from (\ref{ry2}) and (\ref{ry11}), respectively, and
\begin{align} \label{R0}
v= 1 & \!-\!\frac{\alpha}{1440} \left\{\alpha  A^2 \left[m^2 \left(-4636 z^2\!+\!2880 z\!+\!135\right)\!+\!m
   \left(3496 z^2\!-\!3180 z\!-\!270\right)\right. 
 \!-\!76 z^2\!+\!300 z\!+\!135\right]  \nonumber \\ &  \hspace{2ex} \left.
-60 A \left[-9 \alpha 
   D\!+\!m (9 \alpha  D\!+\!50 \alpha  D z\!+\!20 z\!-\!18)
\!-\!2 z (5  \alpha  D\!+\!2)\!+\!18\right]  
\!+\!  540 D (\alpha  D\!-\!4)\right\} 
\end{align}
from (\ref{ry0}). 
Equations (\ref{R2}) and (\ref{R11}) are, in general, not equivalent for arbitrary $z$. However, in both cases when $z=z_{1}$  or $z=z_{2}$, required by (\ref{ry4}) and (\ref{ry31}), they express the same condition. This shows that equations (\ref{ry2}) and (\ref{ry11}) are equivalent, just as are (\ref{ry4}) and (\ref{ry31}), so equations (\ref{ry0})-(\ref{ry31}) supply only three independent conditions.

Solving (\ref{R2})  for $D$ yields
\begin{align} \label{DD2}
D & =  \frac{-27 \alpha  A+27 \alpha  A m-1216 \alpha  A m z^2+708
   \alpha  A m z+608 \alpha  A z^2-444 \alpha  A z-96 z+72}{12
   \alpha  (20 z+3)}.
\end{align}
Substitution of $B$ and $D$ into (\ref{R0}) gives a long formula for the wave's velocity
\begin{align} 
\label{v2}v & = -\left\{\alpha ^2 A^2 \left[24320 (m (1607 m-1382)+323) z^4-3840 (m (7703  m-7368)+1791) z^3 
\right. \right.  \nonumber\\ & \hspace{5ex} \left.
+576 (m (5919 m-7324)+2044) z^2+4320 (m-1) (42 m-19) z+1215 (m-1)^2\right]
\\ & \hspace{5ex} \left.
+69120 \alpha  A (2 m-1) z (2 z (76 z-43)-3)-2880 (8 z (34 z+75)+45)\right\}\; /\; \left[5760 (20 z+3)^2\right] \nonumber
\end{align}
the explicit form of which will be presented in the next section, after specifying the branch of $z$ and taking into account conditions implied by periodicity and volume conservation.

\section{Periodicity and volume conservation conditions}
\label{sss3}

 Denote  
\begin{equation} \label{uu}
u(By,m) = \dn^2(By,m) \pm \sqrt{m}\cn(By,m)\dn(By,m).
\end{equation}
The periodicity condition implies
\begin{equation} \label{lam}
u\left(B L,m\right)=u(0,m) \quad \Longrightarrow \quad  L=\frac{4\,K(m)}{B} ,
\end{equation}
where $K(m)$ is the  complete elliptic integral of the first kind. 
Note that the wavelength $L$ given by (\ref{lam}) is two times greater than that for a single $\cn^2$ periodic solution \cite{IKRR17}.

Then volume conservation  requires
\begin{equation} \label{Vcons}
\int_0^L \eta_{\pm}(By,m)\,dy = \frac{A}{2}\int_0^L u(By,m)\,dy + D\,L =0 .
\end{equation} 
Volume conservation means that elevated and depressed (with respect to the mean level) volumes are the same over the period of the wave.

From properties of elliptic functions
\begin{equation} \label{dj}
\int_0^L u(By,m)\,dy =
\frac{E(\text{am}(4  K(m)|m)|m)}{B} = \frac{4\, E(m)}{B},
\end{equation}
where ~$E(\Theta|m)$ is the elliptic integral of the second kind, $\textrm{am}(x|m)$ is  the Jacobi elliptic function amplitude and $ E(m)$ is the complete elliptic integral of the second kind. 
Then from (\ref{Vcons})-(\ref{dj})  one obtains $D$ in the form
\begin{equation} \label{Dvc}
D=-\frac{A\, E(m)}{2\, K(m)}.
\end{equation}

In order to obtain explicit expressions for coefficients $A,B,D,v$ one has to specify $z$. Choose the positive root first.

{\bf Case ~${\bf z=z_2}=\frac{43+\sqrt{2305}}{152}\approx 0.59875$.}\\[1ex]
With this choice $A>0$ and the cnoidal wave has crests elevation larger than troughs depression with respect to still water level.

Substitution of $z_2$ into equation (\ref{R2}) (or equivalently (\ref{R11})) supplies another relation between $A$ and $D$, giving 
\begin{equation} \label{Dvcz}
D=-\frac{12(-51+\sqrt{2305})+37(5-m) \alpha A}{444\, \alpha  }.
\end{equation}
Equating (\ref{Dvc}) with (\ref{Dvcz}) one obtains ($EK(m)$ is given by (\ref{EKm}))
\begin{equation} \label{Avcz}
A= \frac{12 \left(51-\sqrt{2305}\right)}{37 \alpha \,EK(m)}.
\end{equation}
With this $A$ 
\begin{align} \label{Bz2}
B & = \sqrt{\frac{12(\sqrt{2305}-14)}{703 \beta\, EK(m)}},  \\ \label{Dz2}
D & = - \frac{6 \left(51-\sqrt{2305}\right)}{37 \alpha \,EK(m)} \frac{E(m)}{K(m)}.  
\end{align}

Velocity formula (\ref{v2}) simplifies to
\begin{align}\label{vz2}
v  & = \frac{9439-69 \sqrt{2305}}{5476} +\left(\frac{7811 \sqrt{2305}-377197}{520220}\right)\frac{\left(m^2+14 m+1\right)}{\left[EK(m)\right]^2} \\ & \approx 
1.11875 - 0.00420523\; \frac{\left(m^2+14 m+1\right)}{\left[EK(m)\right]^2} . \nonumber
\end{align}

In general, as stated in previous papers \cite{KRI14,IKRR17,RKI} the KdV2 equation imposes one more condition on coefficients of solutions than KdV. 
Let us discuss obtained results in more detail. Coefficients $A,B,D,v$ are related to the function $EK(m)$. This function is plotted in figure \ref{fAK}.  

\begin{figure}[tbh]
\begin{center} 
\resizebox{0.6\columnwidth}{!}{\includegraphics{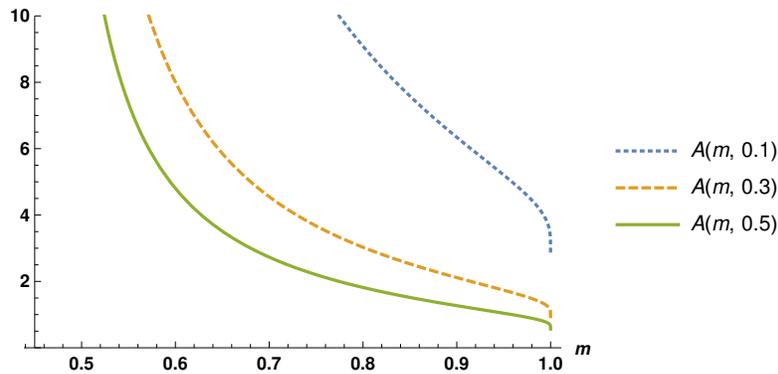}}
\end{center}\vspace{-2mm}
\caption{Amplitude $A(m,\alpha)$ (\ref{Avcz}) of the solution (\ref{etapm}) as function of $m$ for $\alpha=0.1, 0.3, 0.5$.}\label{A(m)z2}
\end{figure}

\begin{figure}[tbh]
\begin{center} 
\resizebox{0.6\columnwidth}{!}{\includegraphics{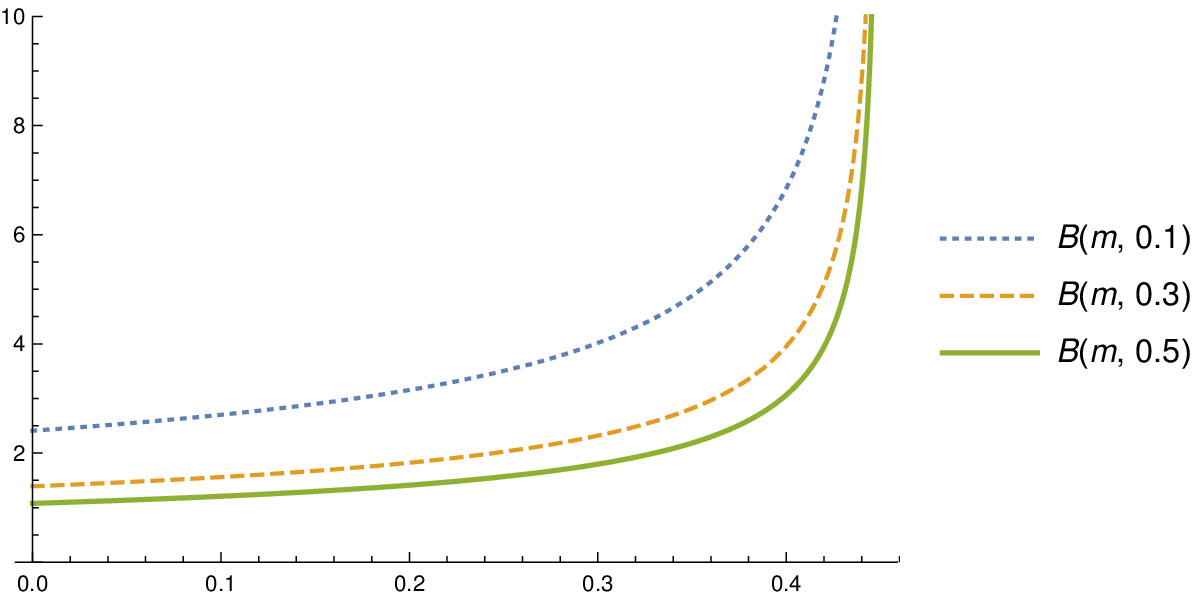}}
\end{center}
\caption{Coefficient $B(m,\beta)$ (\ref{Bz2}) of the solution (\ref{etapm}) as function of $m$ for $\beta=0.1, 0.3, 0.5$.}\label{B(m)z2}
\end{figure}

\begin{figure}[tbh]
\begin{center} 
\resizebox{0.6\columnwidth}{!}{\includegraphics{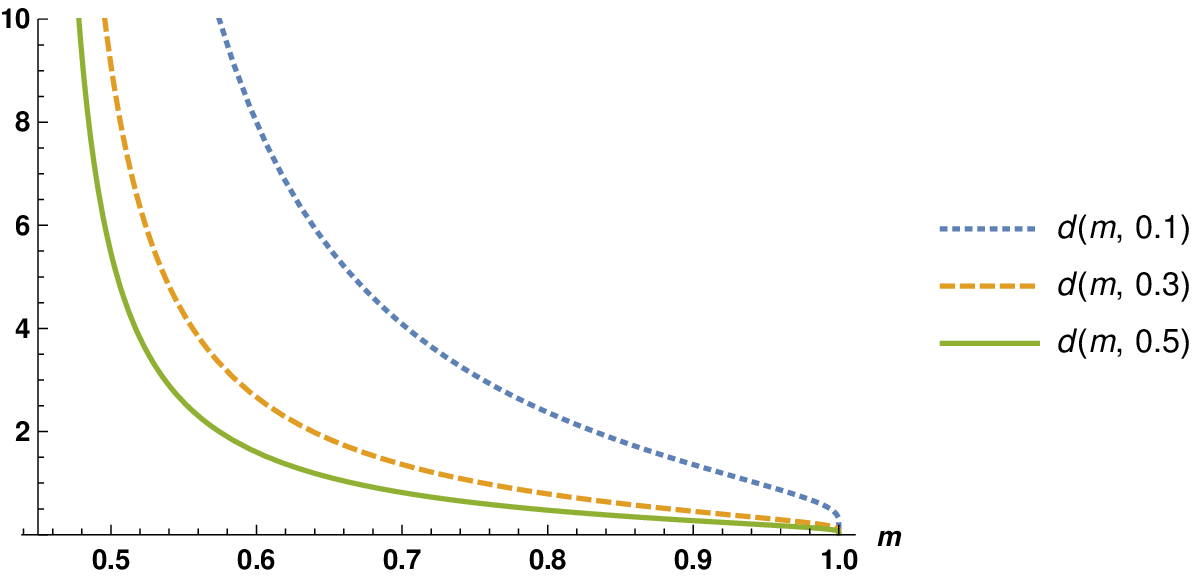}}
\end{center}
\caption{Coefficient $D(m,\alpha)$ (\ref{Dz2}) of the solution (\ref{etapm}) as function of $m$ for $\alpha=0.1, 0.3, 0.5$.}\label{D(m)z2}
\end{figure}

\begin{figure}[tbh]
\begin{center} 
\resizebox{0.5\columnwidth}{!}{\includegraphics{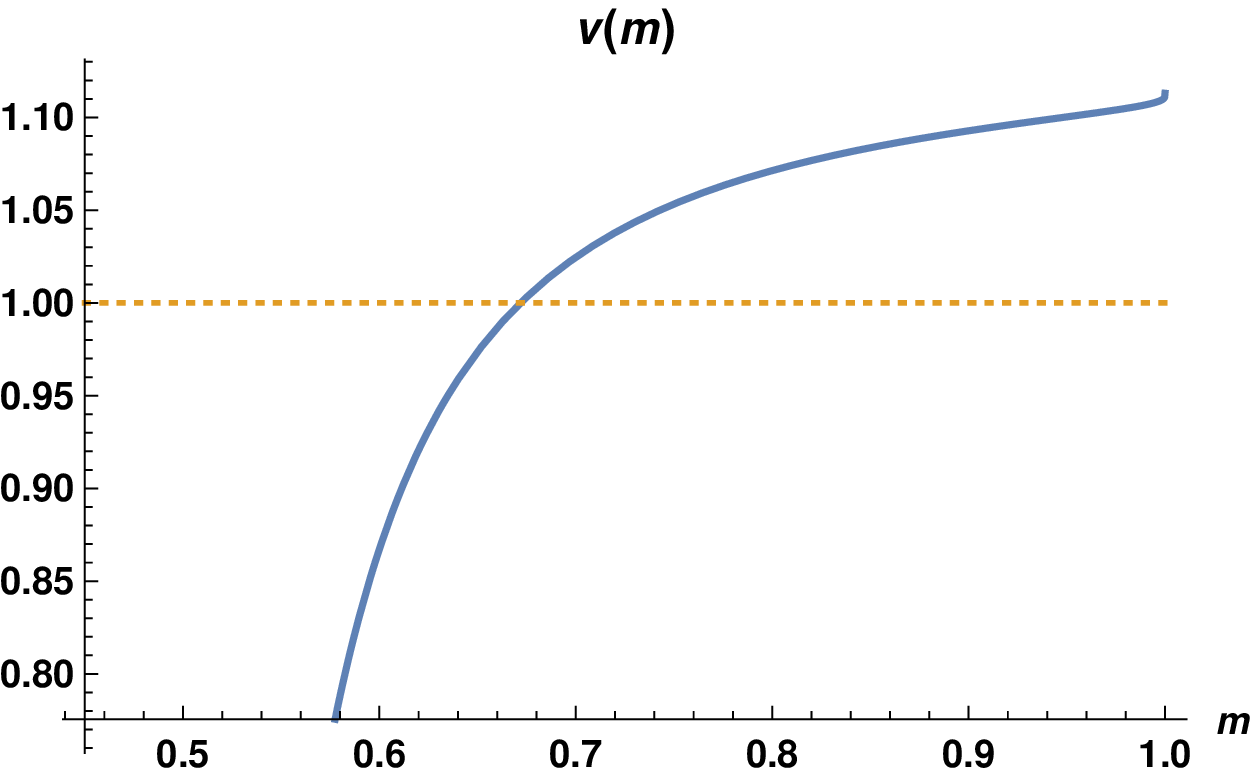}}
\end{center}
\caption{Velocity $v(m)$ (\ref{vz2}) of the solution (\ref{etapm}) as function of $m$.}\label{v(m)z2}
\end{figure}

It is clear that for real-valued $B$ the amplitude $A$ has to be positive, and therefore $m$ must be greater than $\approx 0.45$. 
Since $B$ depends on $m$ this condition imposes a restriction on wavenumbers.
The $m$-dependence of coefficients $A,B,D$ and velocity $v$ (\ref{Avcz})-(\ref{vz2}) are displayed in figures \ref{A(m)z2}-\ref{v(m)z2}. It is worth to note, that $v$ given by (\ref{vz2}) contrary to KdV case (\ref{abdV}) depends only on $m$.

For $m$ close to 1 the wave heigth, that is the difference between the crest's and trough's level is almost equal to $A/2$. 
It is clear from figure \ref{A(m)z2} that the wave height is reasonably small for $m$ close to~1.


{\bf Case} ~${\bf z=z_1}=\frac{43-\sqrt{2305}}{152}\approx -0.0329633$.
\begin{align} \label{Avcz1}
A & =  
\frac{12 \left(\sqrt{2305}+51\right)}{37 \alpha \,EK(m)}, \\ \label{Bz1}
B & = 
\sqrt{-\frac{12(\sqrt{2305}+14)}{703 \beta\, EK(m)}} ,  \\ \label{Dz1}
D & = 
- \frac{6 \left(\sqrt{2305}+51\right)}{37 \alpha \,EK(m)} \frac{E(m)}{K(m)}. 
\end{align}

Velocity formula (\ref{v2}) simplifies to
\begin{align}\label{vv2z1}
v  & = -\frac{9439+69 \sqrt{2305}}{5476} +\left(\frac{7811 \sqrt{2305}+377197}{520220}\right)\frac{\left(m^2+14 m+1\right)}{\left[EK(m)\right]^2}
\\ & \approx 
-2.32866 + 1.44594 \;\frac{\left(m^2+14 m+1\right)}{\left[EK(m)\right]^2} . \nonumber
\end{align}

In this case $B$ is real-valued when $EK(m)$ is negative, that is for $m$ less that $\approx 0.45$ (see, e.g, figure \ref{fAK}). But this means that 
$A$ is negative, that is the cnoidal wave has an inverted shape (crests down, troughs up). The following figures \ref{A(m)z1}-\ref{v(m)z1} illustrate examples of $m$-dependence of coefficients $A,B,D,v$ for 
$m<0.449$.

\begin{figure}[tbh]
\begin{center} 
\resizebox{0.6\columnwidth}{!}{\includegraphics{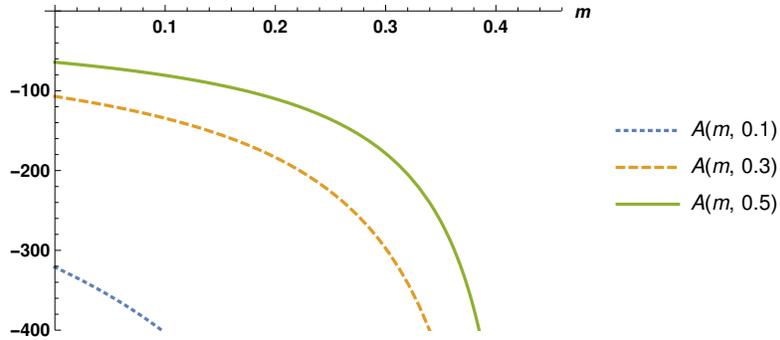}}
\end{center}\vspace{-2mm}
\caption{Amplitude $A(m,\alpha)$ (\ref{Avcz1}) of the solution (\ref{etapm}) as function of $m$ for $\alpha=0.1, 0.3, 0.5$.}\label{A(m)z1}
\end{figure}
\begin{figure}[tbh]
\begin{center} 
\resizebox{0.6\columnwidth}{!}{\includegraphics{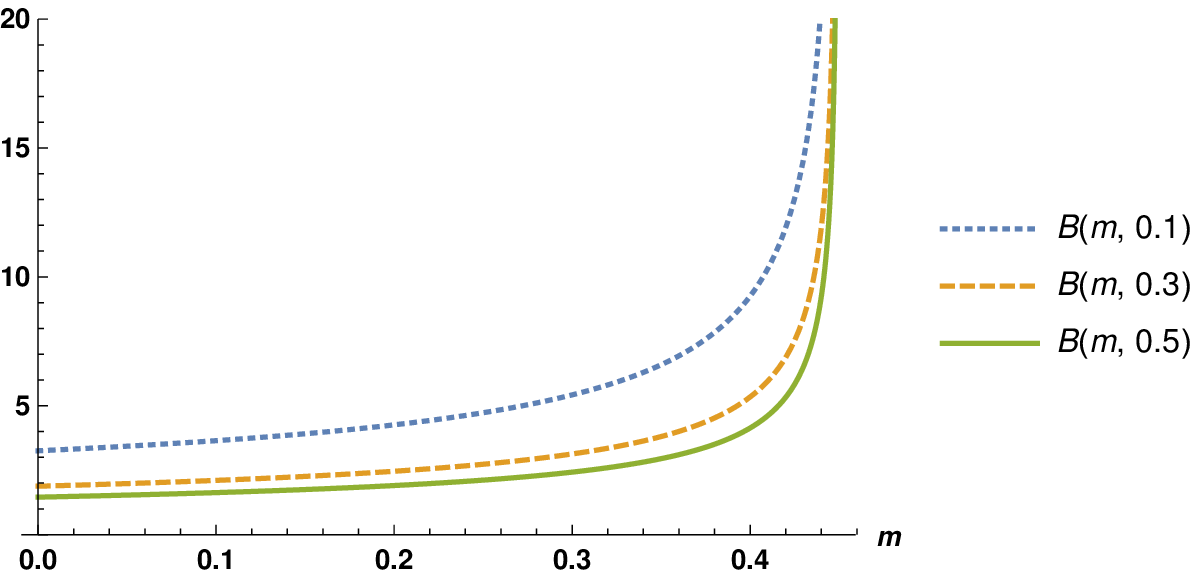}}
\end{center}
\caption{Coefficient $B(m,\beta)$ (\ref{Bz1}) of the solution (\ref{etapm}) as function of $m$ for $\beta=0.1, 0.3, 0.5$.}\label{B(m)z1}
\end{figure}
\begin{figure}[tbh]
\begin{center} 
\resizebox{0.6\columnwidth}{!}{\includegraphics{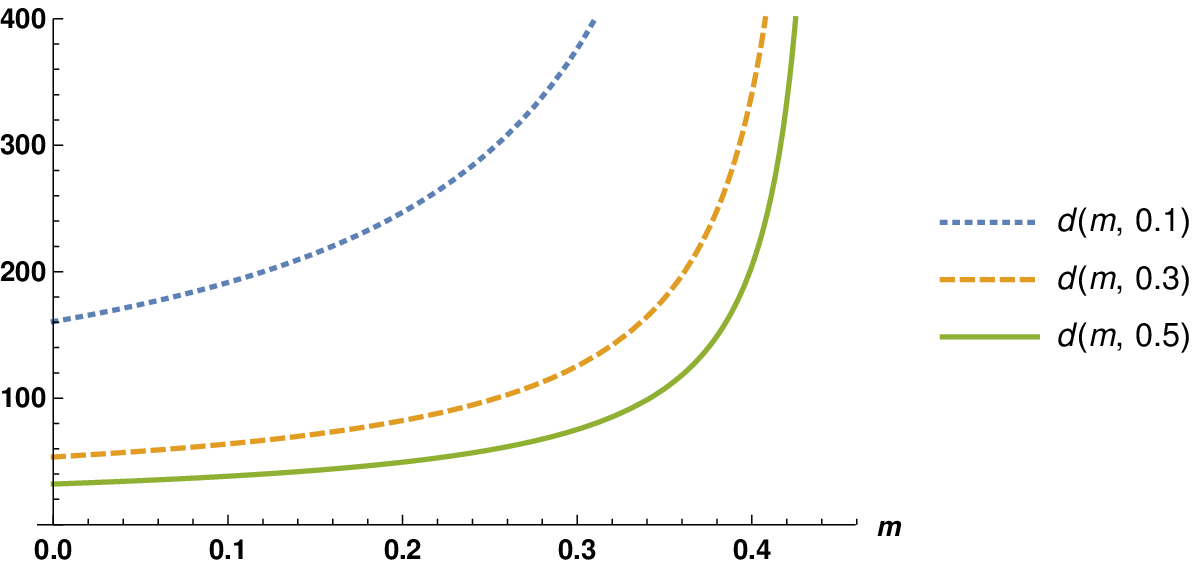}}
\end{center}
\caption{Coefficient $D(m,\alpha)$ (\ref{Dz1}) of the solution (\ref{etapm}) as function of $m$ for $\alpha=0.1, 0.3, 0.5$.}\label{D(m)z1}
\end{figure}

\begin{figure}[tbh]
\begin{center} 
\resizebox{0.5\columnwidth}{!}{\includegraphics{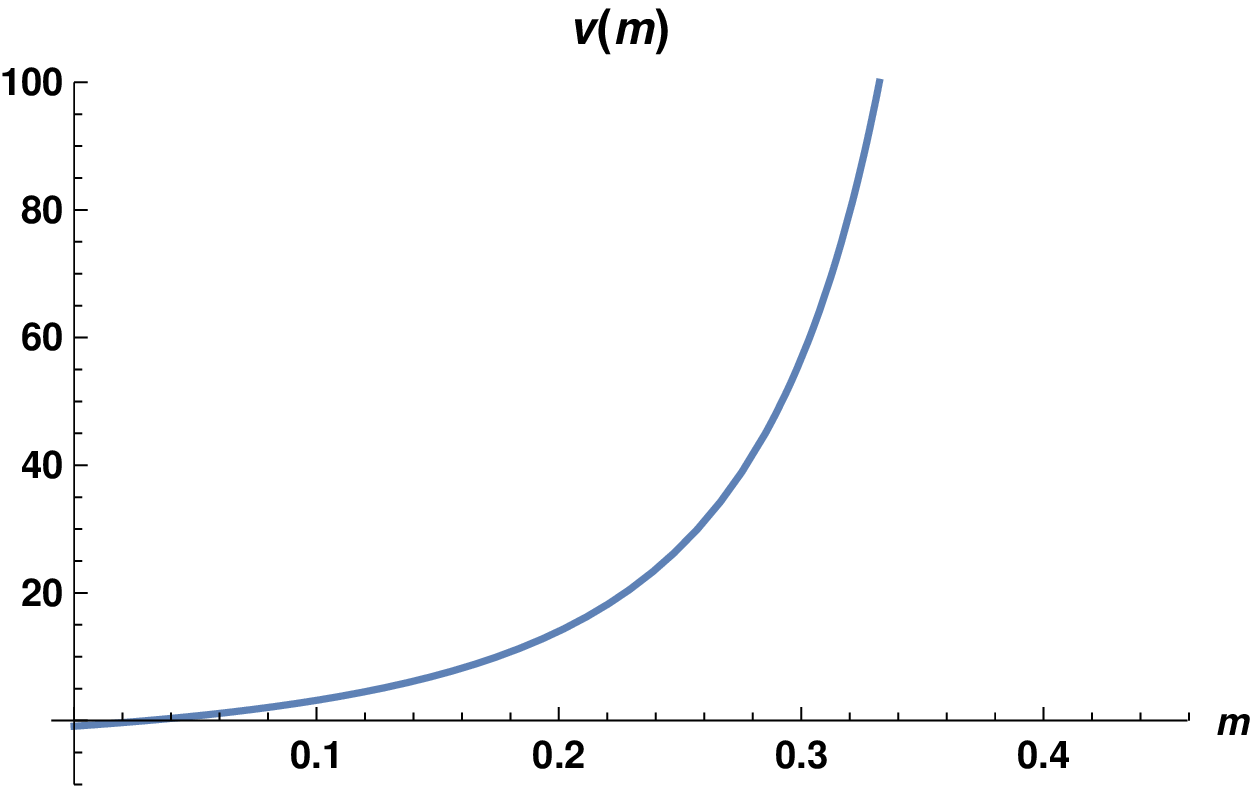}}
\end{center}
\caption{Velocity $v(m)$ (\ref{vv2z1}) of the solution (\ref{etapm}) as function of $m$.}\label{v(m)z1}
\end{figure}

\section{Examples, numerical simulations} \label{sss4}

Table \ref{tab1} contains several examples of coefficients $A,B,D,v$ and the wavelength $L$  of superposition solutions to KdV2 for some particular values of $\alpha,\beta$ and $m$ for the branch $z=z_2$. 

\begin{table}[bth] 
\caption{Examples of values of $A,B,D,v$ and $L$ for ~$z=z_2$~ case.} \label{tab1}
\begin{center}
\begin{tabular}{||l|l|l|l|l|l|l|l||} \hline 
~~~$\alpha$~~~&~~~~$\beta$~~~~&~~~$m$~~~&~~~$A$~~~&~~~~$B$~~~~&~~~~$D$~~~~&~~~~$v$~~~~&~~~~$L$~~~~  \\  \hline 
 0.10 & 0.10 & 0.99 & 4.108 & 1.5683 & -0.5646 & 1.107 & 9.426 \\ \hline
 0.30 & 0.30 & 0.99 & 1.369 & 0.9054 & -0.1882 & 1.107 & 16.33 \\ \hline
 0.50 & 0.50 & 0.99 & 0.822 & 0.7013 & -0.1129 & 1.107 & 21.08 \\ \hline
 0.30 & 0.30 & 0.80 & 3.028 & 1.3465 & -0.7904 & 1.071 & 6.706 \\ \hline
 0.50 & 0.50 & 0.80 & 1.817 & 1.0430 & -0.4743 & 1.071 & 8.657 \\ \hline 
 \end{tabular}
\end{center}
\end{table}  

\begin{figure}[tbh]
\begin{center} 
\resizebox{0.7\columnwidth}{!}{\includegraphics{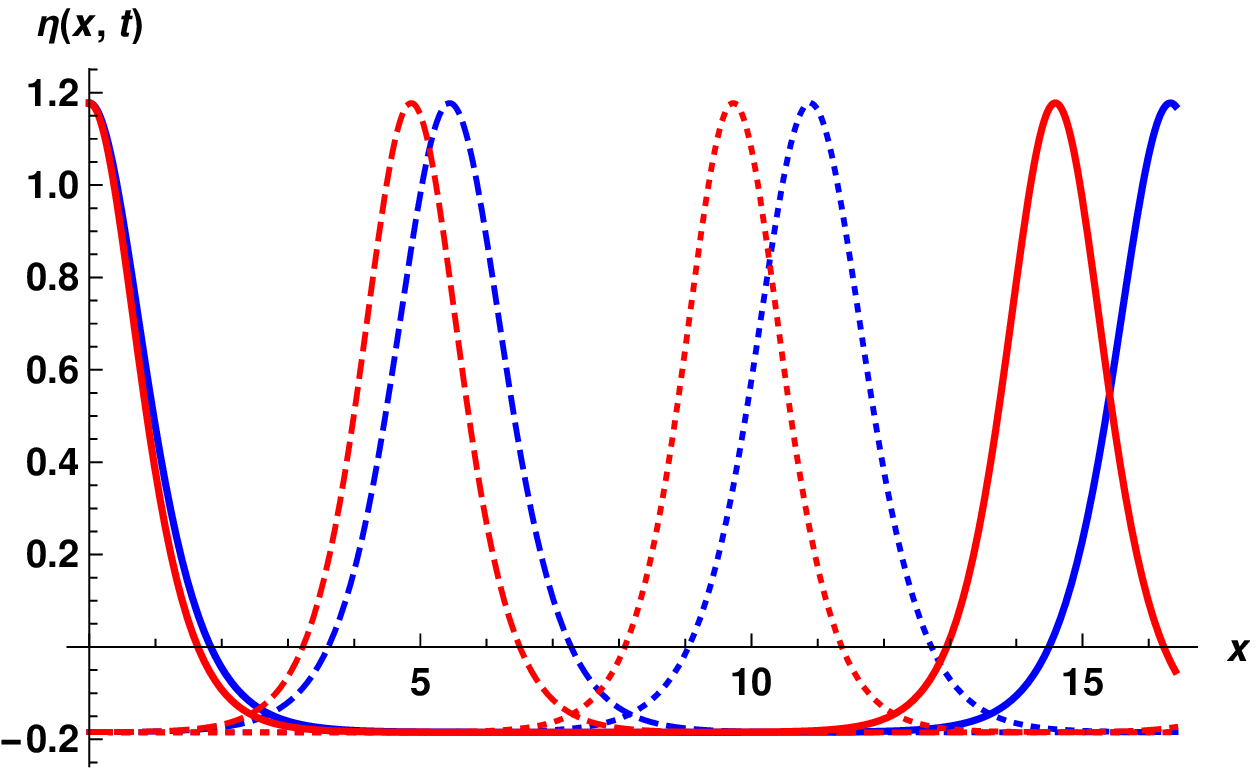}}
\end{center}
\caption{Profiles of the KdV2 (blue lines) and KdV solutions (red lines) for $\alpha=\beta=0.3$ and $m=0.99$. Solid lines correspond to $t=0,T$, dashed lines to $t=T/3$ and dotted lines to $t=2T/3$, respectively ($T$ is the wave period).}\label{kdv12}
\end{figure}

Figure \ref{kdv12} displays a comparison of a solution of KdV2 to solution of KdV. For comparison, parameters $\alpha,\beta$ of the equations 
were chosen to be $\alpha=\beta=0.3$. Compared are waves corresponding to $m=0.99$. Coefficients $A,B,D,v$ of KdV2 solution are given in the second raw of Table \ref{tab1}. For comparison KdV solution is chosen with the same $A$ but $B,D,v$ are given by (\ref{f2-11}), (\ref{d000}) and (\ref{abdV}), respectively. 

Table \ref{tab2} gives two examples of coefficients $A,B,D,v$ and the wavelength $L$  of superposition solutions to KdV2 for some particular values of $\alpha,\beta$ and small $m$ for the branch $z=z_1$. 

\begin{table}[bth] 
\caption{Examples of values of $A,B,D,v$ and $L$ for ~$z=z_1$~ case.} \label{tab2}
\begin{center}
\begin{tabular}{||l|l|l|l|l|l|l|l||} \hline 
~~~$\alpha$~~~&~~~~$\beta$~~~~&~~~$m$~~~&~~~$A$~~~&~~~~$B$~~~~&~~~~$D$~~~~&~~~~$v$~~~~&~~~~$L$~~~~  \\  \hline 
 0.30 & 0.30 & 0.10 & -134.5 & 2.105 & 63.83 & 3.170 & 3.064 \\ \hline
 0.50 & 0.50 & 0.05 & -71.44 & 1.535 & 34.82 & 0.717 & 4.147 \\ \hline
\end{tabular}
\end{center}
\end{table}  

\begin{figure}[tbh]
\begin{center} 
\resizebox{0.7\columnwidth}{!}{\includegraphics{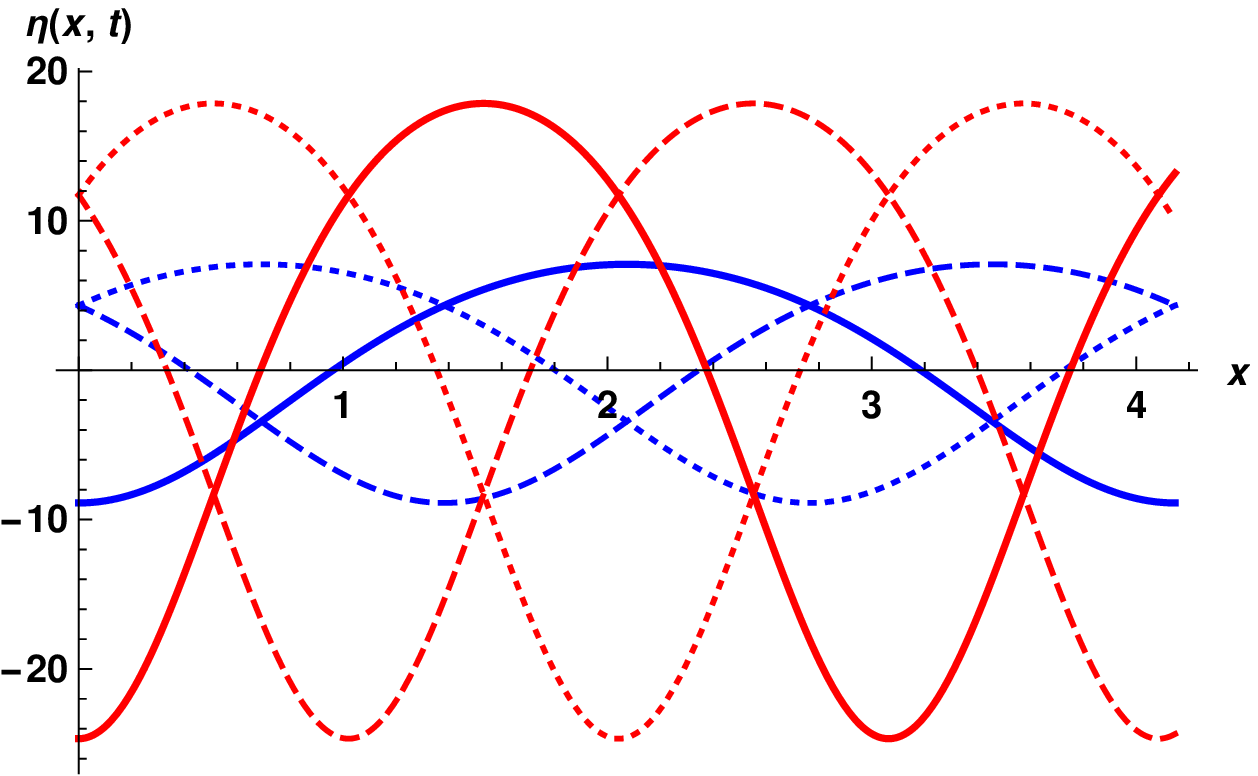}}
\end{center}
\caption{Profiles of the KdV2 solutions for $\alpha=\beta=0.5$ and $m=0.05$ (blue lines) and for $\alpha=\beta=0.3$ and $m=0.1$  (red lines). Solid lines correspond to $t=0,T$, dashed lines to $t=T/3$ and dotted lines to $t=2T/3$, respectively ($T$ is the wave period corresponding to the case).}\label{kdv2z1}
\end{figure}

In figure \ref{kdv2z1} profiles of the solution to KdV2 for the case  $\alpha,\beta=0.5$ and $m=0.05$ are displayed for $t=0, T/3, 2T/3, T$. In this case we obtain an inverted cnoidal shape, with crest depression equals to -8.885 and trough elevation equals to 7.088. 

In the case $\alpha,\beta=0.3$ and $m=0.1$ the corresponding values of crest and trough are -24.67 and 17.85, respectively. 

For $m$ close to 1 the wave height is much smaller than the coefficient $A$ and there exist an interval of small $m$ where the wave height is physically relevant.

\begin{figure}[tbh]
\begin{center} 
\resizebox{0.7\columnwidth}{!}{\includegraphics[angle=270]{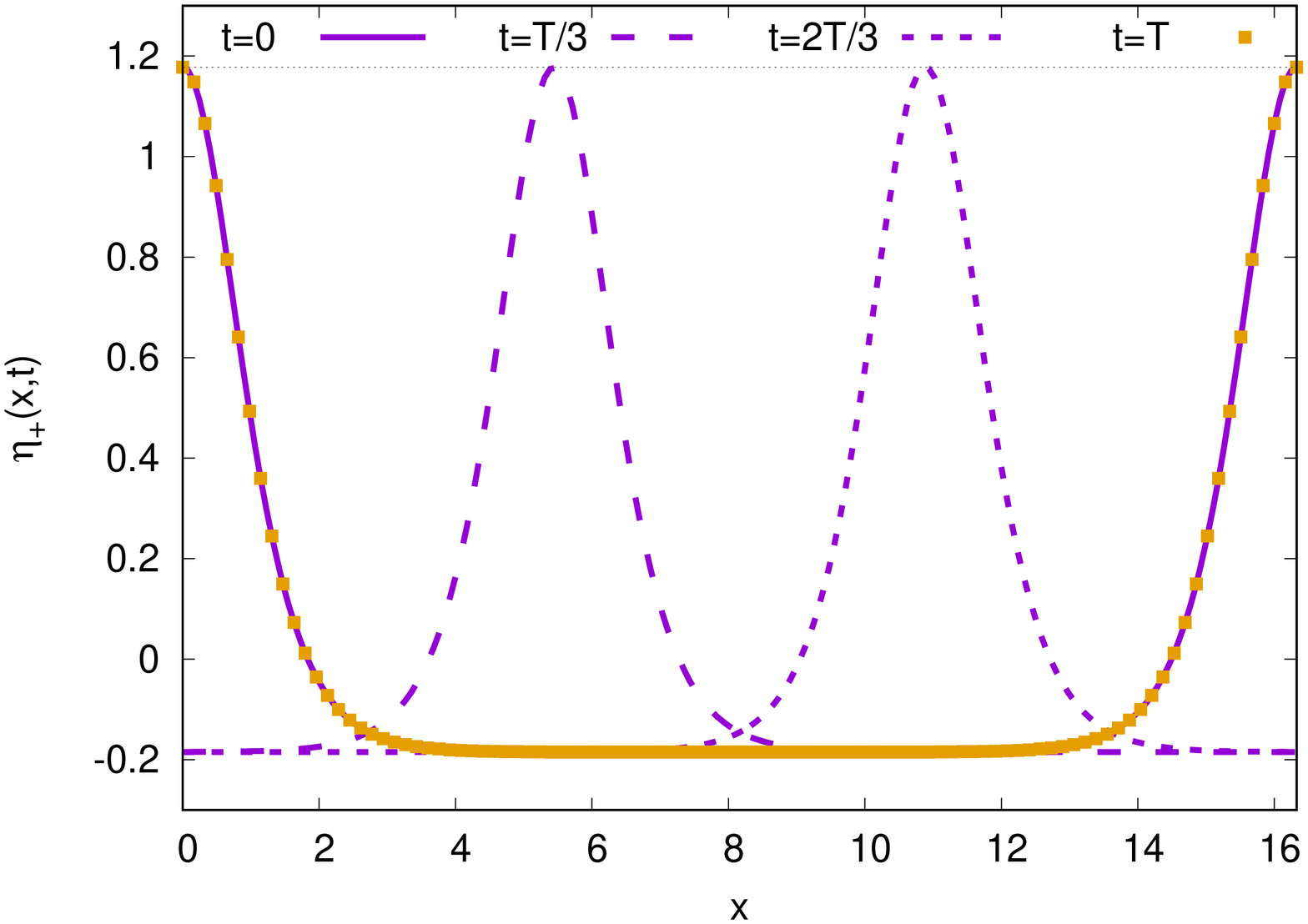}}
\end{center}
\caption{Time evolution of the superposition solution $\eta_+(x,t)$ for $\alpha=\beta=0.3$ and $m=0.99$ obtained in numerical simulations. Profiles of the wave at $t= 0, T/3, 2T/3, T$ are shown, where $T$ is the period.  The $x$ interval is equal to the wavelength.}\label{num_ab003z2}
\end{figure}

Numerical calculations of the time evolution of superposition solutions performed with the finite difference code as used in previous papers \cite{KRR14,KRI14,IKRR17,RKI} confirm the analytic results. Numerical evolution of any of the presented solution shows their uniform motion with perfectly preserved shapes. The case corresponding to $z=z_2$ branch, with parameters listed in the second raw of Table \ref{tab1}, is illustrated in figure \ref{num_ab003z2}. This is the same wave as that displayed in figure \ref{kdv12} (blue lines).

\begin{figure}[tbh]
\begin{center} 
\resizebox{0.7\columnwidth}{!}{\includegraphics[angle=270]{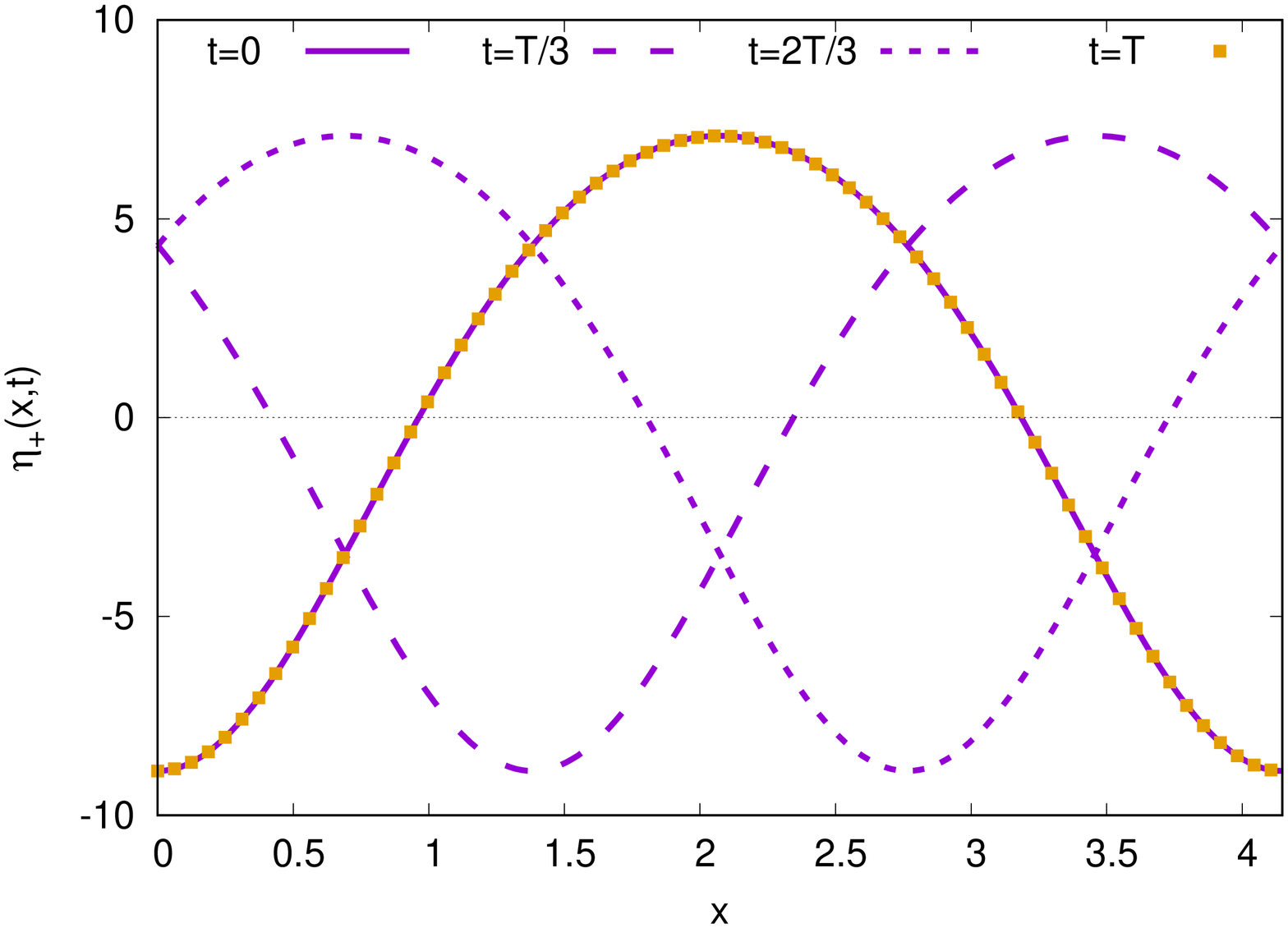}}
\end{center}
\caption{Time evolution of the superposition solution $\eta_+(x,t)$ for $\alpha=\beta=0.3$ and $m=0.05$ obtained in numerical simulations. Profiles of the wave at $t= 0, T/3, 2T/3, T$ are shown, where $T$ is the period.  The $x$ interval is equal to the wavelength.}\label{num_ab003z1}
\end{figure}

The case corresponding to $z=z_1$ branch, with parameters listed in the first raw of Table \ref{tab2}, is illustrated in figure \ref{num_ab003z1}. This is the same wave as that displayed in figure \ref{kdv2z1} (blue lines).

\noindent {\bf Remark:}  \emph{
From periodicity of the Jacobi elliptic functions it follows that 
\begin{equation} \label{pmequiv}
\eta_+(x,t) =\eta_-(x\pm L/2,t).
\end{equation}
This means that both $\eta_{+}(x,t)$ and $\eta_{-}(x,t)$ represent the same wave, but shifted by half of the wavelength with respect to one another.
}

\section{Conclusions} \label{concl}

From the studies on the KdV2 equation presented in this paper and in \cite{KRI14,IKRR17} one can draw the following conclusions.

\begin{itemize}
\item There exist several classes of exact solutions to KdV2 which have the same form as the corresponding solutions to KdV but with slightly different coefficients. These are solitary waves of the form $A\cn^2$ \cite{KRI14}, cnoidal waves $A\cn^2+D$ \cite{IKRR17} and periodic waves in the form (\ref{eypm}), that is, $\frac{A}{2}[\dn^2\pm\sqrt{m}\cn\dn]+D$ studied in the present paper.

\item KdV2 imposes one more condition on coefficients of the exact solutions than KdV.


\item Periodic solutions for KdV2 can appear  in two forms.
The first form,  $A\cn^2 +D$, is, as pointed out in \cite{IKRR17},  physically relevant in two narrow intervals of $m$, one close to $m=0$, another close to $m=1$. 
The second form, given by (\ref{eypm}) gives physically relevant periodic solutions in similar intervals. However, for $m$ close to 1 the superposition solution (\ref{eypm}) forms a wave similar to $A\cn^2 +D$, whereas for small $m$ this wave has inverted cnoidal shape. 

\item All the above mentioned solutions to KdV2 have the same function form as the corresponding KdV solutions but with slightly different coefficients.
\end{itemize}

KdV, besides having single solitonic and periodic solutions, possesses also multi-soliton solutions. The question of whether exact multi-soliton solutions for KdV2 exist is still open. However, numerical simulations presented in the Appendix A, 
in line with the Zabusky-Kruskal numerical experiment \cite{ZK} suggest
such a possibility. A conjecture, that multi-soliton solutions to KdV2 might exist in the same form as KdV  multi-soliton solutions, but with altered coefficients, will be studied soon.

\appendix

\section{Do multi-soliton solutions to KdV2 exist?}
\begin{figure}[tbh]
\begin{center}
\resizebox{0.99\columnwidth}{!}{\includegraphics{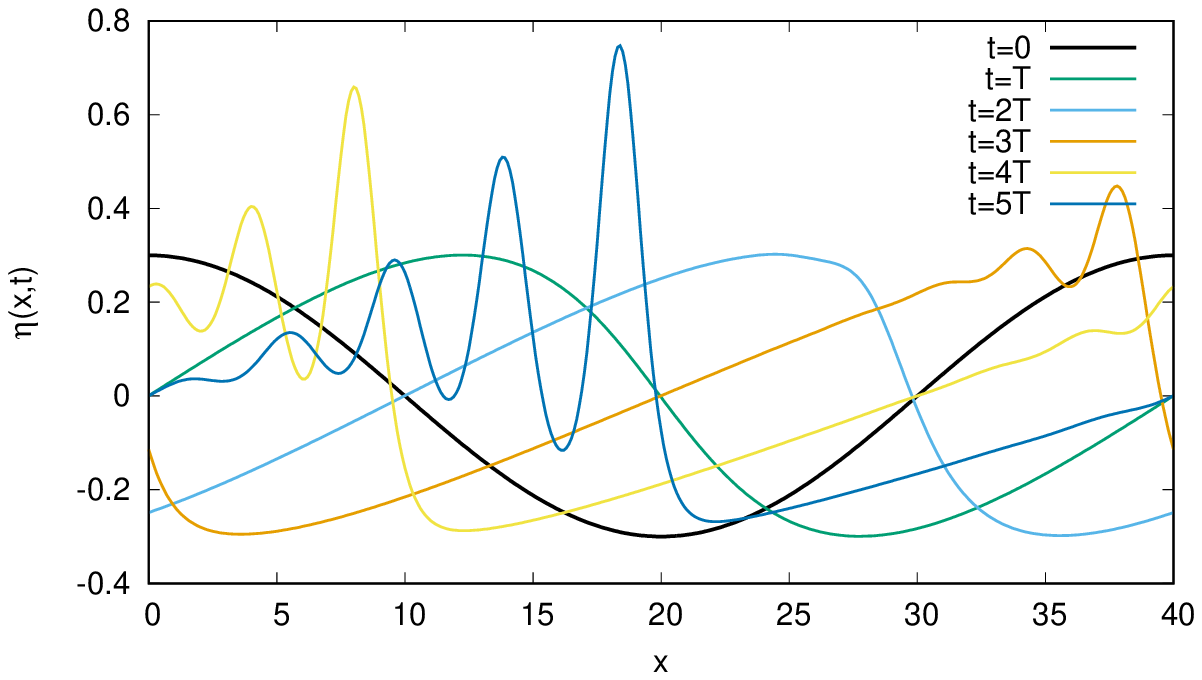}}
\end{center}
\caption{Emergence of soliton trains according to KdV from initial cosine wave. $A=0.3,~T=50$.} \label{c03kdv}
\end{figure}

\begin{figure}[bth]
\begin{center}
\resizebox{0.99\columnwidth}{!}{\includegraphics{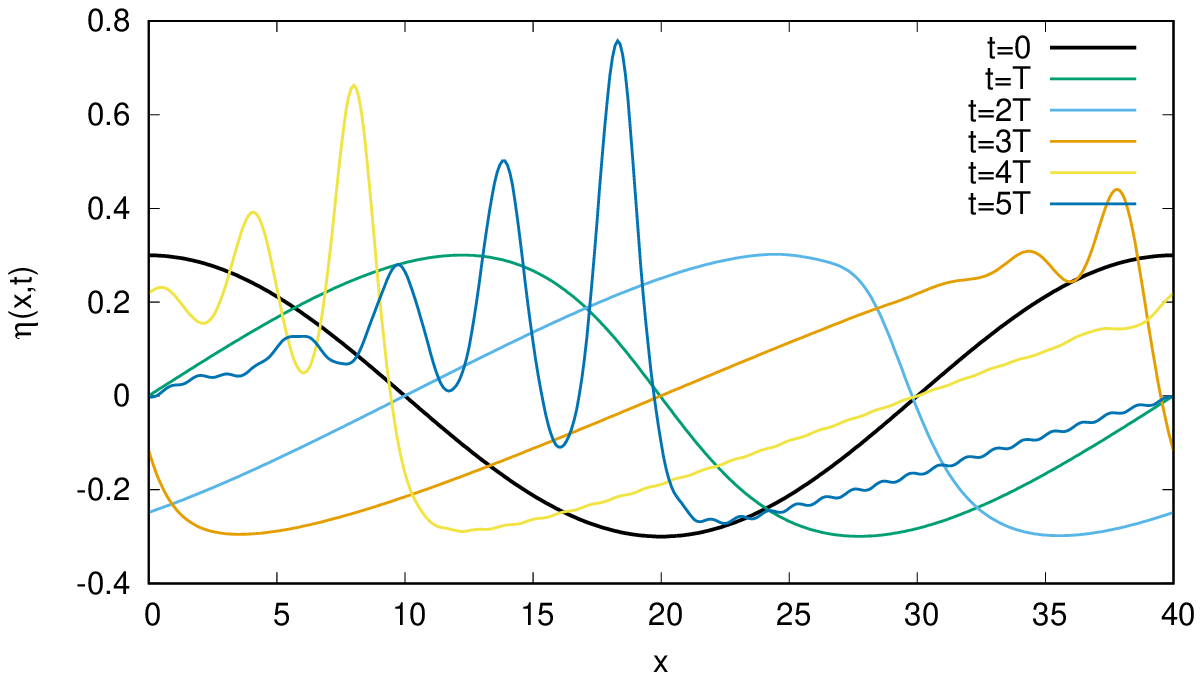}}
\end{center}
\caption{Emergence of soliton trains according to KdV2 from initial cosine wave.  $A=0.3,~T=50$. } \label{c03kdv2}
\end{figure}

 For KdV there exist multi-soliton solutions which can be obtained e.g. using the inverse scattering method \cite{GGKM,AbC} or the Hirota method \cite{Hirota}. The fact that KdV2 is non-integrable would seem to exclude the existence of multi-soliton solutions to KdV2. On the other hand, numerical simulations demonstrate, that for some initial conditions a train of KdV2 solitons, almost the same as of KdV solitons emerges from the cosine wave as in Zabusky and Kruskal \cite{ZK} numerical simulation. We describe such numerical simulation, see figures \ref{c03kdv} and \ref{c03kdv2}.


Initial conditions for both simulations were chosen as a hump $\eta(x,0)=A\cos(\frac{\pi}{40}(x+20))$ for $0\le x\le 40$ and  $\eta(x,0)=0$ for $x>40$ moving to the right. 
Then such a wave was evolved by a finite difference method code developed  in \cite{KRR14,KRI14,KRI15}.  There is a surprising similarity of  trains of solitons obtained in evolutions with KdV and KdV2. This behaviour might suggest the possible existence of multi-soliton KdV2 solutions.

 In a multi-soliton solution of KdV each soliton has a different amplitude. Otherwise these amplitudes are arbitrary. If multi-soliton solutions to KdV2 exist we would expect some restrictions on these amplitudes.


\section*{Conflicts of Interest}
The authors declare that they have no conflicts of interest.

\end{document}